\documentclass[a4paper,12pt,number]{elsarticle}

\usepackage{amssymb,amsmath}
\usepackage{graphicx}
\usepackage{geometry}
\usepackage{xcolor}
\usepackage[normalsize]{subfigure}
\usepackage[colorlinks,citecolor=red,urlcolor=red]{hyperref}

\newcommand{\Eref}[1]{Eq.~\eqref{eq:#1}}
\newcommand{\Sref}[1]{Section~\ref{sec:#1}}
\newcommand{\Fref}[1]{Fig.~\ref{fig:#1}}

\newcommand{\N}{\vek{N}}

\newcommand{\bmath}[1]{{\boldsymbol #1}}
\newcommand{\vek}[1]{\bmath{#1}} 

\newcommand{\je}{\vek{j}}
\newcommand{\se}{L}
\newcommand{\tse}{\vek{\se}}

\newcommand{\ee}{\vek{e}}
\newcommand{\x}{\vek{x}}
\newcommand{\y}{\vek{y}}
\newcommand{\kk}{\vek{k}}
\newcommand{\xk}{\x^\kk}

\newcommand{\m}{\vek{m}}

\newcommand{\LL}{L} 
\newcommand{\tL}{\vek{\LL}}

\newcommand{\PUC}{\mathcal{Y}}
\newcommand{\lPUC}{Y}

\newcommand{\dse}{\delta\vek{\se}}
\newcommand{\G}{\Gamma}
\newcommand{\tG}{\vek{\G}}

\newcommand{\mG}{\mtrx{\Gamma}}

\newcommand\de{\,{\mathrm d}} 

\newcommand{\TP}[1]{#1^\N}
\newcommand{\meas}[1]{\left|#1\right|}

\newcommand{\imu}{\mathrm{i}}

\newcommand{\FT}[1]{\widehat{#1}}

\newcommand{\set}[1]{\mathbb{#1}}
\newcommand{\ZN}{\set{Z}^{\N}}

\newcommand{\mtrx}[1]{\mathsf{#1}}
\newcommand{\me}{\mtrx{e}}

\newcommand{\DFT}{\mtrx{F}}
\newcommand{\iDFT}{\mtrx{F}^{-1}}

\newcommand{\mdse}{\mtrx{\delta\mtrx{L}}}
\newcommand{\trn}{^{\mathsf T}}

\newcommand{\errR}{\eta_{\rm r}}
\newcommand{\errE}{\eta_{\rm e}}

\newcommand{\myMatrix}[2]{%
\left[
\begin{array}{#1}%
#2%
\end{array}%
\right]
}

\newcommand{\ite}[1]{^{(#1)}}
\newcommand{\norm}[1]{\| #1 \|}

\newcommand{\FFTH}{\mathrm{FFTH}}
\newcommand{\CG}{\mathrm{CG}}
\newcommand{\BiCG}{\mathrm{BiCG}}
\newcommand{\BiorCG}{\mathrm{(Bi)CG}}

\journal{Journal of Computational Physics}

\begin{document}
 
\begin{frontmatter}

\title{Accelerating a FFT-based solver for numerical homogenization of
  periodic media by conjugate gradients}

\author[ctu]{Jan Zeman\corref{cor}}
\ead{zemanj@cml.fsv.cvut.cz}
\ead[url]{http://mech.fsv.cvut.cz/\~{}zemanj} 
\cortext[cor]{Corresponding author. Tel.:~+420-2-2435-4482;
fax~+420-2-2431-0775}

\author[ctu]{Jaroslav Vond\v{r}ejc}
\ead{vondrejc@gmail.com}
\address[ctu]{Department of Mechanics, Faculty of Civil
  Engineering, Czech Technical University in Prague,
  Th\' akurova 7, 166 29 Prague 6, Czech Republic}

\author[cideas]{Jan Nov\'{a}k}
\address[cideas]{Centre for Integrated Design of Advanced Structures,
  Faculty of Civil Engineering, Czech Technical University in Prague,
  Th\' akurova 7, 166 29 Prague 6, Czech Republic}
\ead{novakj@cml.fsv.cvut.cz}

\author[ctumath]{Ivo Marek}
\ead{marek@mbox.ms.cuni.cz}
\address[ctumath]{Department of Mathematics, Faculty of Civil
  Engineering, Czech Technical University in Prague,
  Th\' akurova 7, 166 29 Prague 6, Czech Republic}

\begin{abstract}
  In this short note, we present a new technique to accelerate the
  convergence of a FFT-based solver for numerical homogenization of
  complex periodic media proposed by Moulinec and
  Suquet~\cite{Moulinec:1994:FNMC}. The approach proceeds from
  discretization of the governing integral equation by the
  trigonometric collocation method due to
  Vainikko~\cite{Vainikko:2000:FSLS}, to give a linear system which
  can be efficiently solved by conjugate gradient
  methods. Computational experiments confirm robustness of the
  algorithm with respect to its internal parameters and demonstrate
  significant increase of the convergence rate for problems with
  high-contrast coefficients at a low overhead per iteration.
\end{abstract}


\end{frontmatter}

\section{Introduction}\label{sec:introduction}

A majority of computational homogenization algorithms rely on the
solution of the unit cell problem, which concerns the determination of
local fields in a representative sample of a heterogeneous material
under periodic boundary conditions. Currently, the most efficient
numerical solvers of this problem are based on discretization of
integral equations. In the case of particulate composites with smooth
bounded inclusions embedded in a matrix phase, the problem can be
reduced to internal interfaces and solved with remarkable accuracy and
efficiency by the fast multipole method, see~\cite[and references
therein]{Greengard:2006:EH}. An alternative method has been proposed
by Suquet and Moulinec~\cite{Moulinec:1994:FNMC} to treat problems
with general microstructures supplied in the form of digital
images. The algorithm is based on the Neumann series expansion of the
inverse to an operator arising in the associated Lippmann-Schwinger
equation and exploits the Fast Fourier Transform~(FFT) to evaluate the
action of the operator efficiently.

The major disadvantage of the FFT-based method consists in its poor
convergence for composites exhibiting large jumps in material
coefficients. To overcome this difficulty, Eyre and Milton proposed
in~\cite{Eyre:1999:FNS} an accelerated scheme derived from a modified
integral equation treated by means of the series expansion
approach. In addition, Michel et al.~\cite{Michel:2000:CMB} introduced
an equivalent saddle-point formulation solved by the Augmented
Lagrangian method. As clearly demonstrated in a numerical study by
Moulinec and Suquet~\cite{Moulinec:2003:CFFT}, both methods converge
considerably faster than the original variant; the number of
iterations is proportional to the square root of the phase contrast
instead of the linear increase for the basic scheme. However, this
comes at the expense of increased computational cost per iteration and
the sensitivity of the Augmented Lagrangian algorithm to the setting
of its internal parameters.

In this short note, we introduce yet another approach to improve the
convergence of the original FFT-based scheme~\cite{Moulinec:1994:FNMC}
based on the trigonometric collocation method~\cite{Saranen:1996:TCM}
and its application to the Helmholtz equation as introduced by
Vainikko~\cite{Vainikko:2000:FSLS}. We observe that the discretization
results in a system of linear equations with a structured dense
matrix, for which a matrix-vector product can be computed efficiently
using FFT, cf.~\Sref{methodology}. It is then natural to treat the
resulting system by standard iterative solvers, such as the Krylov
subspace methods, instead of the series expansion
technique. In~\Sref{results}, the potential of such approach is
demonstrated by means of a numerical study comparing the performance
of the original scheme and the conjugate- and biconjugate-gradient
methods for two-dimensional scalar electrostatics.

\section{Methodology}\label{sec:methodology}
In this section, we briefly summarize the essential steps of the
trigonometric collocation-based solution to the unit cell problem by
adapting the original exposition by Vainikko~\cite{Vainikko:2000:FSLS}
to the setting of electrical conduction in periodic composites. In
what follows, $a$, $\vek{a}$ and $\vek{A}$ denote scalar, vector and
second-order tensor quantities with Greek subscripts used when
referring to the corresponding components,
e.g. $A_{\alpha\beta}$. Matrices are denoted by a serif font
(e.g. $\mtrx{A}$) and a multi-index notation is employed, in which
$\set{R}^\set{\N}$ with $\N = (N_1, \ldots, N_d )$ represents
$\set{R}^{N_1 \times \cdots \times N_d}$ and $\mtrx{A}^\kk$ stands for
the $(k_1, \ldots, k_d)$-th element of the matrix $\mtrx{A} \in
\set{R}^\N$.

\subsection{Problem setting}
We consider a composite material represented by a periodic unit cell
$\PUC = \prod_{\alpha=1}^d ( -\lPUC_\alpha, \lPUC_\alpha ) \subset
\set{R}^d$. In the context of linear electrostatics, the associated
unit cell problem reads as
\begin{eqnarray}\label{eq:problem_def_1}
  \vek{\nabla} \times \ee( \x ) = \vek{0}, & 
  \vek{\nabla} \cdot \je( \x ) = \vek{0},  & 
  \je( \x ) = \tse( \x ) \cdot \ee( \x ), \;
  \x \in \PUC
\end{eqnarray}
where $\ee$ is a $\PUC$-periodic vectorial electric field, $\je$
denotes the corresponding vector of electric current and $\tse$ is a
second-order positive-definite tensor of electric conductivity. In
addition, the field $\ee$ is subject to a constraint
\begin{equation}\label{eq:problem_def_2}
  \ee^0 = \frac{1}{\meas{\PUC}} \int_\PUC \ee( \x ) \de\x,
\end{equation}
where $\ee^0$ denotes a prescribed macroscopic electric field and
$\meas{\PUC}$ represents the $d$-dimensional measure of $\PUC$.

Next, we introduce a homogeneous reference medium with constant
conductivity $\tse^0$, leading to a decomposition of the electric
current field in the form
\begin{eqnarray}\label{eq:current_decomp}
  \je(\x) = \tL^0 \cdot \ee(\x) + \dse(\x) \cdot \ee(\x),
&&
  \dse(\x) = \tL(\x) - \tL^0.
\end{eqnarray}
The original
problem~\eqref{eq:problem_def_1}--\eqref{eq:problem_def_2} is then
equivalent to the periodic Lippmann-Schwinger integral equation,
formally written as
\begin{eqnarray}\label{eq:LS_real}
  \ee( \x ) 
  +
  \int_\PUC 
  \tG^0( \x - \y ) 
  \cdot 
  \Bigl( 
    \delta\tL( \y ) \cdot \ee( \y ) 
  \Bigr)
  \de \vek{y}
  = 
  \ee^0,
  &&
  \x \in \PUC,
\end{eqnarray}
where the $\tG^0$ operator is derived from the Green's function of the
problem~\eqref{eq:problem_def_1}--\eqref{eq:problem_def_2} with $\tL(
\x ) = \tL^0$ and $\ee^0 = \vek{0}$. Making use of the convolution
theorem, \Eref{LS_real} attains a local form in the Fourier space:
\begin{equation}\label{eq:LS_Fourier}
  \FT{\ee}( \kk ) 
  =
  \left\{
    \begin{array}{rl}
    \meas{\PUC}^{\frac{1}{2}}\ee^0, & \kk = \vek{0}, \\
    \displaystyle
    - \FT{\tG}^0( \kk ) 
    \cdot
    \FT{ \left( \delta\tL \cdot \ee \right) }( \kk ),
    & \kk \in \set{Z}^d \backslash \{ \vek{0} \},
    \end{array} 
  \right.
\end{equation}
where $\FT{f}( \kk )$ denotes the Fourier coefficient of $f( \x )$ for
the $\kk$-th frequency given by
\begin{eqnarray}\label{eq:fourier_coeff_def}
  \FT{f}( \kk ) 
  = 
  \int_\PUC f( \x ) \varphi_{-\kk} ( \x ) 
  \de\x, &&
  \varphi_{ \kk }( \vek{x} ) 
  = 
  | \PUC |^{-\frac{1}{2}}
  \exp\left( 
    \imu \pi \sum_{\alpha=1}^d \frac{x_\alpha k_\alpha}{\lPUC_\alpha} 
  \right),
\end{eqnarray}
"$\imu$'' is the imaginary unit and
\begin{equation}
  \FT{\tG}^0( \kk ) 
  =
  \left\{
    \begin{array}{rl}
      \vek{0}, & \kk = \vek{0}, \\
      \displaystyle
      \frac{\kk \otimes \kk}{\kk \cdot \tL^0 \cdot \kk}, & 
      \kk \in \set{Z}^d \backslash \{ \vek{0} \},
    \end{array}
  \right. 
\end{equation}
Here, we refer to~\cite{Eyre:1999:FNS,Vinogradov:2008:AFFT} for
additional details.

\subsection{Discretization via trigonometric collocation}

Numerical solution of the Lippmann-Schwinger equation is based on a
discretization of a unit cell $\PUC$ into a regular periodic grid with
$N_1 \times \cdots \times N_d$ nodal points and grid spacings $\vek{h}
= ( 2\lPUC_1/N_1, \ldots, 2 \lPUC_d/N_d )$. The searched field $\ee$
in \eqref{eq:LS_real} is approximated by a trigonometric polynomial
$\TP{\ee}$ in the form~(cf.~\cite{Vainikko:2000:FSLS})
\begin{eqnarray}\label{eq:trig_pol_def}
  \ee( \x ) 
  \approx 
  \TP{\ee}( \x ) 
  = 
  \sum_{\vek{\kk} \in \ZN}
  \FT{\ee}( \kk ) \varphi_{\kk}( \x ), 
  && 
  \x \in \PUC,
\end{eqnarray}
where $\N = ( N_1, \ldots, N_d )$, $\FT{\ee}$ designates the Fourier
coefficients defined in~\eqref{eq:fourier_coeff_def} and
\begin{equation}
  \ZN 
  = 
  \left\{ 
    \vek{k} \in \set{Z}^d : 
    -\frac{N_\alpha}{2} < k_\alpha \leq \frac{N_\alpha}{2}, 
    \alpha = 1, \ldots, d 
  \right\}.
\end{equation}

We recall, e.g. from~\cite{Vainikko:2000:FSLS}, that the $\alpha$-th
component of the trigonometric polynomial expansion $\TP{e_\alpha}$
admits two equivalent finite-dimensional representations. The first
one is based on a matrix $\FT\me_\alpha \in \set{C}^{\N}$ of the
Fourier coefficients of the $\alpha$-th component and
equation~\eqref{eq:trig_pol_def} with $\FT{e}_\alpha( \kk ) =
\FT\me_\alpha^\kk$. Second, the data can be entirely determined by
interpolation of nodal values
\begin{eqnarray}\label{eq:nodal_representation}
  \TP{e}_\alpha( \x ) 
  =
  \sum_{\kk \in \ZN}
  \me_\alpha^{\kk} \TP{\varphi}_{\kk}( \vek{x} ), &&
  \alpha = 1, \ldots, d
\end{eqnarray}
where $\me_\alpha \in \set{R}^\N$ is a matrix storing electric field
values at grid points, $\me_\alpha^{\kk} = \TP{e}_\alpha( \xk )$ is
the corresponding value at the $\kk$-th node with coordinates $\xk = (
k_1 h_1, \ldots, k_dh_d )$ and basis functions
\begin{equation}\label{eq:ahoj}
  \varphi^\N_{\kk}( \x )
  =
  | \N |^{-1}
  \sum_{\m \in \ZN}
  \exp 
  \left\{  
    \imu \pi  \sum_{\alpha=1}^{d} 
    m_\alpha \left( 
      \frac{x_\alpha}{\lPUC_\alpha} - \frac{2k_\alpha}{N_\alpha} 
    \right) 
  \right\}
\end{equation}
satisfy the Dirac delta property $\varphi^\N_{\kk}( \x^\m ) =
\delta_{\m \kk}$ with $\meas{\N} = \prod_{\alpha=1}^d N_\alpha$.  Both
representations can be directly related to each other by
\begin{eqnarray}\label{eq:DFT_def}
  \FT{\me}_\alpha = \DFT \me_\alpha, 
  && 
  \me_\alpha = \iDFT \FT{\me}_\alpha,
\end{eqnarray}
where the Vandermonde matrices $\DFT \in \set{C}^{\N\times \N}$ and
$\iDFT \in \set{C}^{\N \times \N}$
\begin{eqnarray}
  \DFT^{\kk\m} 
  & = &
  | \PUC |^{-\frac{1}{2}}
  \exp   \left( -
    \sum_{\alpha=1}^{d} 
    2 \pi \imu \frac{k_\alpha m_\alpha}{N_\alpha}
  \right),
  \\
  \left( \iDFT \right)^{\kk\m}
  & = &
  | \PUC |^{\frac{1}{2}}
  | \N |^{-1} 
  \exp \left( 
    \sum_{\alpha=1}^{d}    
    2 \pi \imu \frac{k_\alpha m_\alpha}{N_\alpha}
  \right),
\end{eqnarray}
implement the forward and inverse Fourier transform, respectively,
e.g.~\cite[Section~4.6]{Golub:1996:MC}.

The trigonometric collocation method is based on the projection of the
Lippmann-Schwinger equation~\eqref{eq:LS_real} to the space of the
trigonometric polynomials of the form $\left\{\sum_{\kk \in \ZN} c_\kk
  \varphi_\kk, c_\kk \in \set{C} \right\}$,
cf.~\cite{Saranen:1996:TCM,Vainikko:2000:FSLS}. In view
of~\Eref{nodal_representation}, this is equivalent to the collocation
at grid points, with the action of $\tG^0$ operator evaluated from the
Fourier space expression~\eqref{eq:LS_Fourier} converted to the nodal
representation by~\eqref{eq:DFT_def}$_2$. The resulting system of
collocation equations reads
\begin{eqnarray}\label{eq:linear_system}
\left( \mtrx{I} + \mtrx{B} \right)  \mtrx{e} = \mtrx{e}^0, 
\end{eqnarray}
where $\me \in \set{R}^{d \times \N}$ and $\me^0 \in \set{R}^{d \times
  \N}$ store the corresponding solution and of the macroscopic field,
respectively. Furthermore, $\mtrx{I}$ is the $d \times d \times \N
\times \N$ unit matrix and the non-symmetric matrix $\mtrx{B}$ can be
expressed, for the two-dimensional setting, in the partitioned format
as
\begin{equation}\label{eq:mtrx_B}
\mtrx{B}
=
 \myMatrix{cc}{%
   \iDFT & \mtrx{0} \\
   \mtrx{0} & \iDFT 
 }
 \myMatrix{cc}{%
   \FT{\mG}^0_{11} & \FT{\mG}^0_{12} \\
   \FT{\mG}^0_{21} & \FT{\mG}^0_{22} 
 }
 \myMatrix{cc}{%
   \DFT & \mtrx{0} \\
   \mtrx{0} & \DFT 
 }
 \myMatrix{cc}{%
 \mdse_{11} & \mdse_{12} \\
 \mdse_{21} & \mdse_{22}
},
\end{equation}
with an obvious generalization to an arbitrary dimension. Here,
$\FT{\mG}^0_{\alpha\beta} \in \set{R}^{\N \times \N}$ and
$\mdse_{\alpha \beta} \in \set{R}^{\N \times \N}$ are diagonal
matrices storing the corresponding grid values, for which it holds
\begin{eqnarray}
  \left(\FT{\mG}^0_{\alpha\beta}\right)^{\kk\kk} 
  = 
  \FT{\G}^0_{\alpha\beta}(\kk), &
  \displaystyle
  \mdse^{\kk\kk}_{\alpha\beta} = \delta\LL_{\alpha\beta}(\x^\kk), &
  \alpha, \beta = 1, \ldots, d \mbox{ and } \kk \in \ZN.
\end{eqnarray}

\subsection{Iterative solution of collocation equations}
%
It follows from~\Eref{mtrx_B} that the cost of the multiplication by
$\mtrx{B}$ or by $\mtrx{B}\trn$ is driven by the forward and inverse
Fourier transforms, which can be performed in $O(\meas{\N} \log
\meas{\N} )$ operations by FFT techniques. This makes the resulting
system~\eqref{eq:linear_system} ideally suited for iterative solvers.

In particular, the original Fast Fourier Transform-based
Homogenization~(FFTH) scheme formulated by Moulinec and Suquet
in~\cite{Moulinec:1994:FNMC} is based on the Neumann expansion of the
matrix inverse $(\mtrx{I} + \mtrx{B})^{-1}$, so as to yield the $m$-th
iterate in the form
\begin{equation}\label{eq:orig_fft}
\mtrx{e}\ite{m} 
=
\sum_{j=0}^{m} 
\left( -\mtrx{B} \right)^j
\mtrx{e}^0.
\end{equation}
Convergence of the series~\eqref{eq:orig_fft} was comprehensively
studied in~\cite{Eyre:1999:FNS,Vinogradov:2008:AFFT}, where it was
shown that the optimal rate of convergence is achieved for
\begin{equation}\label{eq:opt_ref}
\tL^0 = \frac{\lambda_{\min} + \lambda_{\max}}{2} \vek{I}, 
\end{equation}
with $\lambda_{\min}$ and $\lambda_{\max}$ denoting the minimum and
maximum eigenvalues of $\tL( \x )$ on~$\PUC$ and $\vek{I}$ being the
identity tensor.

Here, we propose to solve the non-symmetric
system~\eqref{eq:linear_system} by well-established Krylov subspace
methods, in particular, exploiting the classical Conjugate
Gradient~($\CG$) method~\cite{Hestenes:1952:MCG} and the biconjugate
gradient~($\BiCG$) algorithm~\cite{Fletcher:1976:BCG}. Even though
that $\CG$ algorithm is generally applicable to symmetric and
positive-definite systems only, its convergence in the one-dimensional
setting has been proven by
Vond\v{r}ejc~\cite[Section~6.2]{Vondrejc:2009:AHM}. A successful
application of $\CG$ method to a generalized Eshelby inhomogeneity
problem has also been recently reported by
Nov\'{a}k~\cite{Novak:2008:CESS} and Kanaun~\cite{Kanaun:2009:FAEF}.

\section{Results}\label{sec:results}
To assess the performance of the conjugate gradient algorithms, we
consider a model problem of the transverse electric conduction in a
square array of identical circular particles with $50\%$ volume
fraction.  A uniform macroscopic field $\ee^0 = (1,0)$ is imposed on
the corresponding single-particle unit cell, discretized by $\N =
(255, 255)$ nodes\footnote{%
  Note that the odd number of discretization points is used to
  eliminate artificial high-frequency oscillations of the solution in
  the Fourier space, as reported in~\cite[Section
  2.4]{Moulinec:1998:NMC}.}
and the phases are considered to be isotropic with the conductivities
set to $\tse = \vek{I}$ for the matrix phase and to
$\tse=\varrho\vek{I}$ for the particle.

The conductivity of the homogeneous reference medium is parameterized
as
\begin{equation}\label{eq:ref_media_choice}
  \tse^0 ( \omega ) 
  = 
  \left( 1 - \omega + \varrho \omega \right) \vek{I},
\end{equation}
where $\omega = 0.5$ corresponds to the optimal convergence of $\FFTH$
algorithm~\eqref{eq:opt_ref}. All conjugate gradient-related results
have been obtained using the implementations according
to~\cite{Saad:2003:IMSL} and referred to as Algorithm~6.18~($\CG$
method) and Algorithm~7.3~($\BiCG$ scheme). Two termination criteria
are considered. The first one is defined for the $m$-th iteration
as~\cite{Moulinec:1998:NMC}
\begin{equation}\label{eq:errE_def}
  \left( \errE\ite{m} \right)^2
  = 
  \frac{
    \sum_{\kk \in \ZN}
    \left( \kk \cdot \FT{\mtrx{j}}^{\kk(m)} \right)^2
  }{
    \norm{\FT{\mtrx{j}}^{\vek{0}(m)}}^2_2
  } \leq \varepsilon^2,
\end{equation}
and provides the test of the equilibrium
condition~\eqref{eq:problem_def_1}$_2$ in the Fourier space. An
alternative expression, related to the standard residual norm for
iterative solvers, has been proposed by Vinogradov and Milton
in~\cite{Vinogradov:2008:AFFT} and admits the form
\begin{equation}\label{eq:errR_def}
  \errR\ite{m} 
  =
  \frac{
    \norm{\mtrx{L}^0 
      \left( 
        \me\ite{m+1} - \me\ite{m} \right) 
    }_2}{\norm{\me^0}_2} \leq \varepsilon,
\end{equation}
with the additional $\mtrx{L}^0$ term ensuring the proportionality
to~\eqref{eq:errE_def} at convergence. From the numerical
  point of view, the latter criterion is more efficient than the
  equilibrium variant, which requires additional operations per
  iteration. From the theoretical point of view, its usage is
  justified only when supported by a convergence result for the
  iterative algorithm. In the opposite case, the equilibrium norm
  appears to be more appropriate, in order to avoid spurious
  non-physical solutions.

\subsection{Choice of reference medium and norm}
Since no results for the optimal choice of the reference medium are
known for $\BiorCG$-based solvers, we first estimate their sensitivity
to this aspect numerically. The results appear in \Fref{example1a},
plotting the relative number of iterations for $\CG$ and $\BiCG$
solvers against the conductivity of the reference medium parameterized
by $\omega$, recall~\Eref{ref_media_choice}.

\begin{figure}[ht]
\centering
\subfigure[]{\includegraphics[scale=1.05]{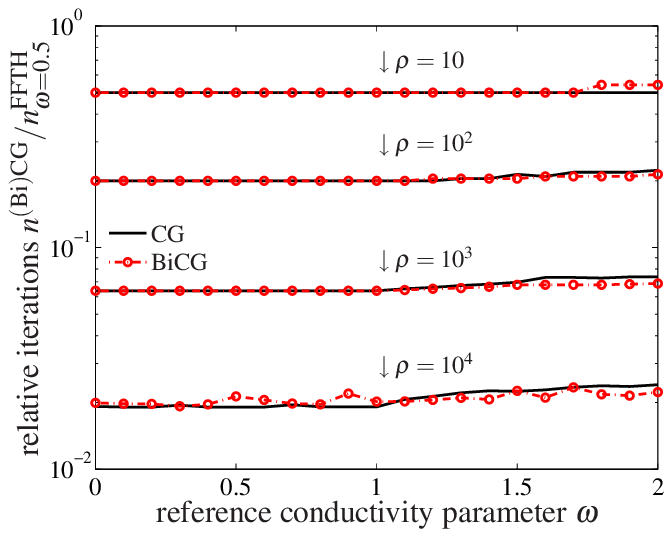}
\label{fig:example1a}
}
~
\subfigure[]{\includegraphics[scale=1.05]{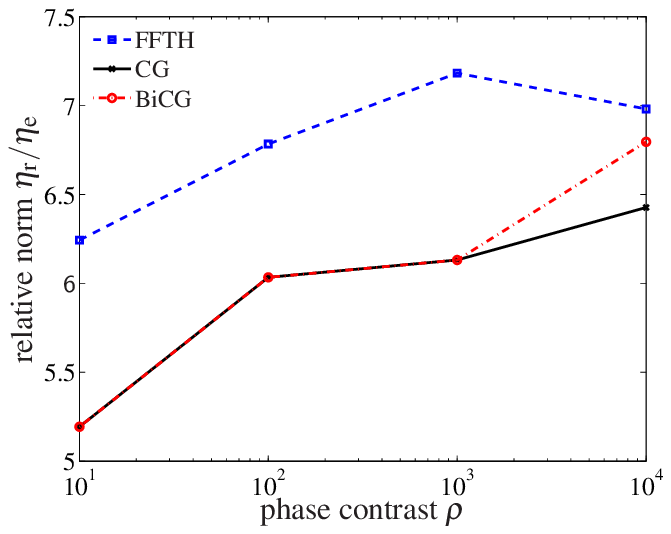}%
\label{fig:example1b}}
\caption{(a)~Relative number of iterations as a function of the
  reference medium parameter $\omega$ and (b)~ratio between residual-
  and equilibrium-based norms at convergence for $\errR$ termination
  condition with tolerance $\varepsilon = 10^{-4}$.}
\label{fig:example1}
\end{figure}

As expected, both $\CG$ and $\BiCG$ solvers achieve a significant
improvement over $\FFTH$ method in terms of the number of iterations,
ranging from $50\%$ for a mildly-contrasted composite down to $2\%$
for $\varrho = 10^4$. Moreover, contrary to all other available
methods, the number of iterations is almost independent of the choice
of the reference medium. We also observe, in agreement with results
by~\cite[Section~6.2]{Vondrejc:2009:AHM} for the one-dimensional
setting, that $\CG$ and $\BiCG$ algorithms generate identical
sequences of iterates; the minor differences visible for $\omega > 1$
or $\varrho = 10^4$ can be therefore attributed to accumulation of
round-off errors. These conclusions hold for both equilibrium- and
residual-based norms, which appear to be roughly proportional for the
considered range of the phase contrasts,
cf.~\Fref{example1b}. Therefore, the residual
  criterion~\eqref{eq:errR_def} will mostly be used in what follows.

\begin{figure}[ht]
\centering
\includegraphics{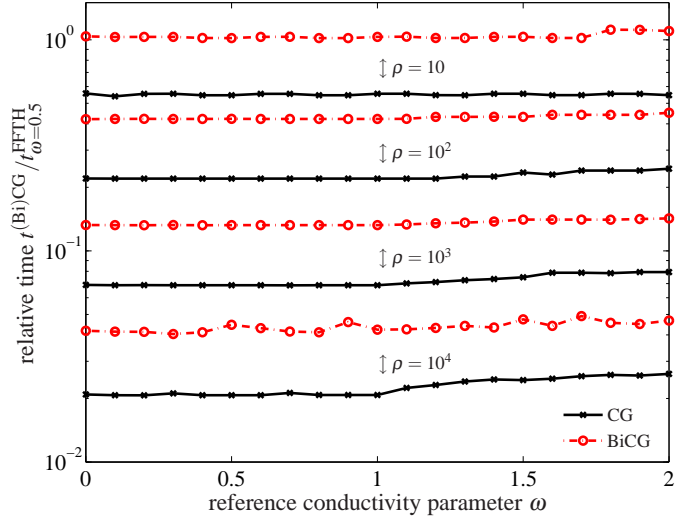}
\caption{Relative CPU time $t$ of $\CG$ and $\BiCG$ solvers plotted
  against the conductivity parameter $\omega$ for $\errR$-based
  termination condition with tolerance $\varepsilon = 10^{-4}$.}
\label{fig:example1c}
\end{figure}

In~\Fref{example1c}, we supplement the comparison by considering the
total CPU time required to achieve a convergence. The data indicate
that the cost of one iteration is governed by the matrix-vector
multiplication, recall~\Eref{mtrx_B}: the overhead of $\CG$ scheme is
about $10\%$ with respect to $\FFTH$~method, while the application of
$\BiCG$ algorithm, which involves $\mtrx{B}$ and $\mtrx{B}\trn$
products per iteration~\cite{Fletcher:1976:BCG}, is about twice as
demanding. As a result, CG algorithm significantly reduces the overall
computational time in the whole range of contrasts, whereas a similar
effect has been reported for the candidate schemes only for $\varrho
\geq 10^3$, cf.~\cite{Moulinec:2003:CFFT}.

\subsection{Influence of phase contrast}
As confirmed by all previous works, the phase contrast $\varrho$~is
the critical parameter influencing the convergence of FFT-based
iterative solvers. In~\Fref{example2}, we compare the scaling of the
total number of iterations with respect to phase contrast for CG and
$\FFTH$ methods, respectively. The results clearly show that the
number of iterations grows as $\sqrt{\varrho}$ instead of the linear
increase for $\FFTH$ method.  This follows from error bounds
\begin{eqnarray}\label{eq:conv_rate}
  \errR\ite{m} \leq \gamma^m \errR\ite{0}, & 
  \displaystyle
  \gamma^\FFTH = \frac{\varrho - 1}{\varrho + 1}, &
  \gamma^\CG = \frac{\sqrt{\varrho}-1}{\sqrt{\varrho}+1}.
\end{eqnarray}
The first estimate was proven in~\cite{Eyre:1999:FNS}, whereas the
second expression is a direct consequence of the condition number of
matrix $\mtrx{B}$ being proportional to $\varrho$ and a well-known
result for the conjugate gradient method, e.g.~\cite[Section
6.11.3]{Saad:2003:IMSL}. The CG-based method, however, failed
  to converge for the infinite contrast limit. Such behavior is
  equivalent to the Eyre-Milton scheme~\cite{Eyre:1999:FNS}. It is,
  however, inferior to the Augmented Lagrangian algorithm, for which
  the convergence rate improves with increasing $\rho$ and the method
  converges even as $\rho \rightarrow \infty$. Nonetheless, such
  results are obtained for optimal, but not always straightforward,
  choice of the parameters~\cite{Michel:2000:CMB}.

\begin{figure}[ht]
\centering
\includegraphics{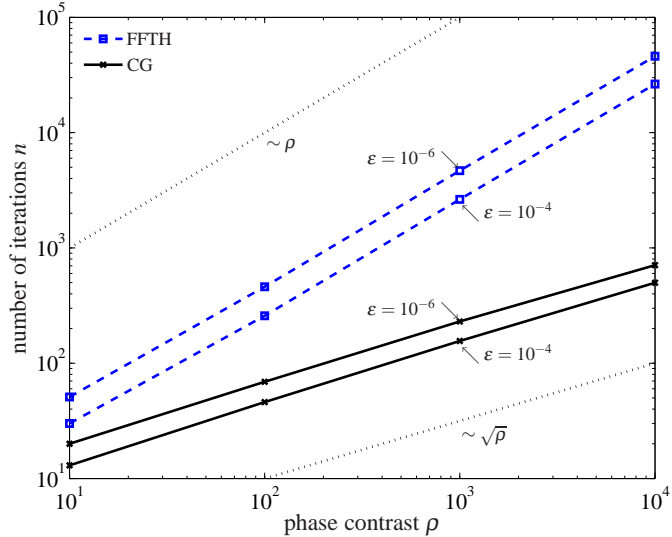}
\caption{Total number of iterations $n$ plotted against phase contrast
  $\varrho\,$; the reference medium corresponds to for $\omega = 0.5$
  and tolerance~$\varepsilon$ is related to $\errR$ norm.}
\label{fig:example2}
\end{figure}

\subsection{Convergence progress}

The final illustration of the $\CG$-based algorithm is provided
by~\Fref{example3}, displaying a detailed convergence behavior for
both low- and high-contrast cases. The results in~\Fref{example3a}
correspond well with estimates~\eqref{eq:conv_rate} for both residual
and equilibrium-based norms.  Influence of a higher phase contrast is
visible from~\Fref{example3b}, plotted in the full logarithmic
scale. For $\FFTH$ algorithm, two regimes can be clearly
distinguished. In the first few iterations, the residual error rapidly
decreases, but the iterates tend to deviate from equilibrium. Then,
both residuals are simultaneously reduced. For $\CG$ scheme, the
increase of the equilibrium residual appears only in the first
iteration and then the method rapidly converges to the correct
solution. However, its convergence curve is irregular and the
algorithm repeatedly stagnates in two consecutive iterations. Further
analysis of this phenomenon remains a subject of future work.

\begin{figure}[ht]
\centering
\subfigure[]{\includegraphics[scale=1.05]{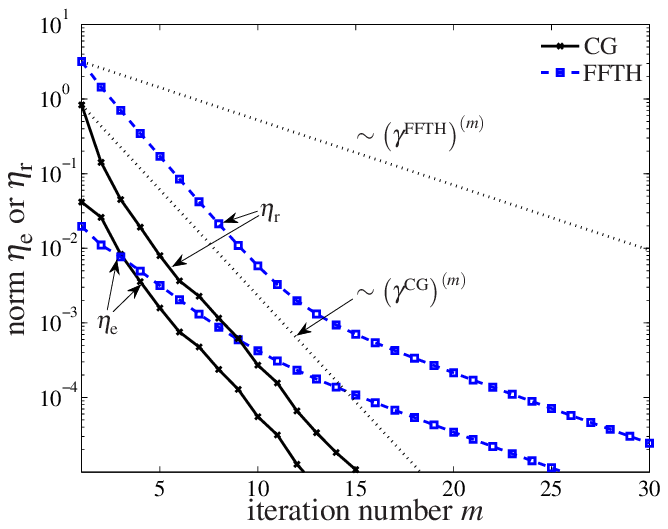}%
\label{fig:example3a}
}
~
\subfigure[]{\includegraphics[scale=1.05]{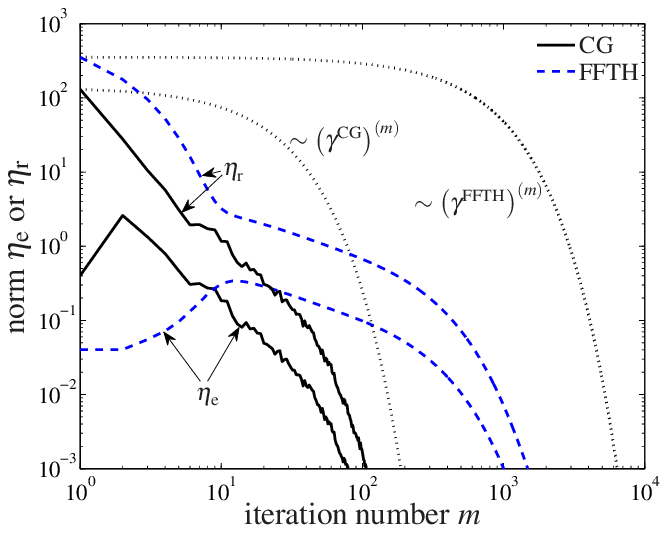}%
\label{fig:example3b}}
\caption{Convergence progress of CG and FFTH methods for (a)~$\varrho
  = 10^1$ and (b)~$\varrho = 10^3$ as quantified by $\errE$ and
  $\errR$ norms; reference medium corresponds to $\omega = 0.5$ and
  the dot-and-dahsed curves indicate the convergence
  rates~\eqref{eq:conv_rate}. }
\label{fig:example3}
\end{figure}

\section{Conclusions}

In this short note, we have presented a conjugate gradient-based
acceleration of the FFT-based homogenization solver originally
proposed by Moulinec and Suquet~\cite{Moulinec:1994:FNMC} and
illustrated its performance on a problem of electric conduction in a
periodic two-phase composite with isotropic phases. On the basis of
obtained results, we conjecture that:

\begin{itemize}

\item[i)] the non-symmetric system of linear
  equations~\eqref{eq:linear_system}, arising from discretization by
  the trigonometric collocation method~\cite{Vainikko:2000:FSLS}, can
  be solved using the standard conjugate gradient algorithm,

\item[ii)] the convergence rate of the method is proportional to the
  square root of the phase contrast,

\item[iii)] the methods fails to converge in the infinite contrast
  limit,

\item[iv)] contrary to available improvements of the original
  FFT-solver~\cite{Eyre:1999:FNS,Michel:2000:CMB}, the cost of one
  iteration remains comparable to the basic scheme and the method is
  insensitive to the choice of auxiliary reference medium.

\end{itemize}

The presented computational experiments provide the first step towards
further improvements of the method, including a rigorous analysis of
its convergence properties, acceleration by multi-grid solvers and
preconditioning and the extension to non-linear problems.

\paragraph{Acknowledgements}
The authors thank Milan Jir\'{a}sek~(Czech Technical University in
Prague) and Christopher Quince~(University of Glasgow) for helpful
comments on the manuscript. This research was supported by the Czech
Science Foundation, through projects No.~GA\v{C}R 103/09/1748,
No.~GA\v{C}R 103/09/P490 and No.~GA\v{C}R 201/09/1544, and by the
Grant Agency of the Czech Technical University in Prague through
project No.~SGS~OHK1-064/10.

\end{document}